# Relaxation Time Approximation for the Wigner-Boltzmann Transport Equation

Aldo R. Fernandes Nt.

Centro Federal de Educação Tecnológica –CEFET/RJ, CEP 20.271-110, Rio de Janeiro, Brazil

aldo.neto@cefet-rj.br

A quasi-distribution function in phase space (based on Wigner functions) is used to write down the quantum version of Boltzmann equation (Wigner-Boltzmann transport equation). The relaxation time approximation is show to be a good approach when defects are homogeneously distributed and linear response to the external electric field is assumed. An expression for relaxation times based on this formalism is deduced, which consider the dependence in the quantum numbers when confinement effects are important (nanowires and nanosheets).

## INTRODUCTION

The Boltzmann equation consists in a powerful way to describe systems which is out of thermodynamic equilibrium [1,2]. Generally, it involves the diffusion or transport of mass, energy or charge that makes the system evolve towards its equilibrium configuration, providing a suitable formalism to calculate a number of properties of a given physical system, such as thermal and electrical conductivities, diffusion coefficients, etc... That makes it a major tool for statistical physics, whose applicability ranges from the microscopic scale up to cosmological domains.

The equation is usually solved for a function, $f(\vec{r}, \vec{p})$, which gives the number of particles of a given system contained in a certain differential volume, $(\vec{r} + d\vec{r}, \vec{p} + d\vec{p})$, of the phase space, consisting, therefore, in a distribution function. Such definition, however, would not be a formal approach if quantum systems are considered, due to the uncertainty principle, which forbids the localization of a particle in a point of the phase space.

Meanwhile, a number of formulations of quantum mechanics in the phase space has been presented in the literature [3], which provides a straightforward way to establish a link between quantum and classical descriptions of nature, associating a quantum state to a function of position and momentum. The most widely known formulation of phase space quantum mechanics is the Weyl-Wigner representation [3-9]. This formulation is quite useful to describe quantum transport process [10] which is important for condensed matter physics [4] and to the understanding of transition for classical statistical physics [5].

This work is divided into four parts. In the first one, the Wigner-Boltzmann equation, i.e., an equation describing the evolution of a quantum phase space "quasi-distribution" function [11,12] is stablished, and the meaning of this function is defined.

In the second part, the formalism is applied to electrical transport in a diffusive media which is subjected to an external electric field. At first, it is demonstrated that, by assuming a linear response of the stationary estate density distribution to the external field, a relaxation time appears naturally in the calculations when collisions are due to homogeneously distributed scattering centres. An expression for relaxation times as a function of quantum numbers is so deduced.

The relaxation time approximation plays an important role in a number of methods to solve the Boltzmann equation, and also in more general ones involving fluid dynamics such as lattice Boltzmann methods [13,14]. Following its broad applicability, their limits of validity have been widely studied also [15,16].

The results are presented in the third part. It is demonstrated that even when considering transport due to the electron at the Fermi surface, the confinement effect in structures such as nanowires and nanosheets can make a constant relaxation time approach unrealistic in some cases, so that the dependence on confinement quantum numbers should play an important hole.

The fourth part is devoted to the final conclusions.

# I. ELECTRONIC TRANSPORT IN THE PHASE SPACE

*The Wigner-Boltzmann Equation*

We can formulate the electronic transport by starting from an ideal system consisting in a non-interacting electron gas in a perfect crystalline potential whose solution is known. Effects of imperfections and impurities in the crystal lattice, as well as interaction with phonons and between the electrons are represented by collisions. A potential $V_{ext}(\vec{r},t)$ represents the external field.

In the case of a nanowire where electrons are confined in transversal directions and free to move continuously along the axial direction, in the conduction energy band, they can be described by the plane waves solution [17,18] in which the crystalline potential is replaced by the 2D confinement potential, $V_{conf}(\vec{r}_\perp)$, and the real electron mass $m$ is replaced by the effective mass $m^*$ [19]. Despite the fact that $V_{conf}(\vec{r}_\perp)$ can have a complicated shape, approaches based on solvable potentials can be used for practical calculations, by considering the suitable conditions [20,21].

The Wigner-Boltzmann equation for electronic transport inside the nanowire is

$$\frac{1}{m^*}\vec{p}\cdot\vec{\nabla}_r f(\vec{r},\vec{p})$$

$$+\frac{1}{i\hbar}\left[V_{conf}\left(\vec{r}_\perp+\frac{i\hbar}{2}\vec{\nabla}_{p_\perp}\right)-V_{conf}\left(\vec{r}_\perp-\frac{i\hbar}{2}\vec{\nabla}_{p_\perp}\right)+V_{ext}\left(\vec{r}+\frac{i\hbar}{2}\vec{\nabla}_p,t\right)-V_{ext}\left(\vec{r}-\frac{i\hbar}{2}\vec{\nabla}_p,t\right)\right]f(\vec{r},\vec{p}) \quad (1)$$

$$+\frac{\partial f(\vec{r},\vec{p})}{\partial t}=-\left(\frac{\partial f(\vec{r},\vec{p})}{\partial t}\right)_{coll}$$

where operatorial functions are defined by the suitable Taylor series expansion, considering the fact that the gradient operators in the function argument on the left acts only on the function variables on the right [8,9].

We shall keep in mind that the function $f(\vec{r},\vec{p})$ in Eq. (1) is not a conventional distribution. If we integrate this solution over the entire momentum space, we obtains the particle distribution $\rho(\vec{r})$ in the configuration space. Meanwhile, if we integrate the solution over the entire configuration space, we obtains the particle distribution $\pi(\vec{p})$ in the momentum space. Therefore, integration over the entire phase space results in the total particle number $N$ inside the system.

These operations can be summarised as follows:

$$\int d\vec{p}\, f(\vec{r},\vec{p})=\rho(\vec{r}), \quad (2a)$$

$$\int d\vec{r}\, f(\vec{r},\vec{p})=\pi(\vec{p}), \quad (2b)$$

$$\int d\vec{r}d\vec{p}\, f(\vec{r},\vec{p})=N. \quad (2c)$$

This definition makes clear the correspondence between the formulation of quantum mechanics in the phase space – using Wigner functions – and the classical Boltzmann theory for the phase space distribution which describes particle systems.

# I. NANOWIRE SBJECTED TO A STATIC ELECTRIC FIELD

*The Wigner-Boltzmann Equation for Diffusive Nanowires*

Let us consider the external potential as a static uniform electric field, applied in the axial $z$ direction, such that $\vec{E}=E\hat{z}$ (if the electric field has non axial components, it can be considered in the confinement potential without loss of generality).

Moreover, we can use state transitions rates to write down the collision term, so that there is no explicit dependency of $f(\vec{r}, \vec{p})$ on time (stationary transport: $\partial f(\vec{r}, \vec{p})/\partial t = 0$). In this case, $V_{ext}(z) = eEz$, where we set the electron charge as $-e$, and Eq. (1) becomes

$$\frac{1}{m^*} \vec{p} \cdot \vec{\nabla}_r f(\vec{r}, \vec{p})$$

$$+ \frac{1}{i\hbar}\left[V_{conf}\left(\vec{r}_\perp + \frac{i\hbar}{2}\vec{\nabla}_{p_\perp}\right) - V_{conf}\left(\vec{r}_\perp - \frac{i\hbar}{2}\vec{\nabla}_{p_\perp}\right)\right] f(\vec{r}, \vec{p}) + eE \frac{\partial f(\vec{r}, \vec{p})}{\partial p_z} \qquad (3)$$

$$= -\left(\frac{\partial f(\vec{r}, \vec{p})}{\partial t}\right)_{coll}$$

where $(\vec{r}_\perp, \vec{p}_\perp) = (x, y, p_x, p_y)$.

Next, we shall found a solution for Eq. (3). Firstly, the equilibrium solution, $f_0(\vec{r}, \vec{p})$, should be stablished, which corresponds to the ideal electron gas,

$$\frac{1}{m^*}\vec{p} \cdot \vec{\nabla}_r f_0(\vec{r}, \vec{p}) + \frac{1}{i\hbar}\left[V_{conf}\left(\vec{r}_\perp + \frac{i\hbar}{2}\vec{\nabla}_{p_\perp}\right) - V_{conf}\left(\vec{r}_\perp - \frac{i\hbar}{2}\vec{\nabla}_{p_\perp}\right)\right] f_0(\vec{r}, \vec{p}) = 0. \qquad (4)$$

The solution can be separated into a transversal part, corresponding to the 2D confined system, and an axial one, corresponding to the plane waves solution (which, in the phase space representation, is a Dirac delta function):

$$f_0(\vec{r}, \vec{p}) = \sum_{\{n_\perp\}, k_z} w_{\{n_\perp\}}(\vec{r}_\perp, \vec{p}_\perp) \eta \delta(p_z - \hbar k_z) n_{0\{n_\perp\}, k_z}, \qquad (5)$$

where $\{n_\perp\}$ is the set of transversal confinement quantum numbers, $k_z$ is the axial wave number and $\eta$ is a normalization factor. The Wigner function $w_{\{n_\perp\}}(\vec{r}_\perp, \vec{p}_\perp)$ [3,6] corresponds to the confinement solution, and the $n_{0\{n_\perp\}, k_z}$ occupation number corresponds to the Fermi-Dirac distribution.

Moreover, by considering the length of the nanowire as being sufficiently large to be set as infinite, we can treat $k_z$ as a continuous variable, so that Eq. (5) becomes

$$f_0(\vec{r}, \vec{p}) = \sum_{\{n_\perp\}} w_{\{n_\perp\}}(\vec{r}_\perp, \vec{p}_\perp) \int \frac{dk_z}{2\pi} \delta(p_z - \hbar k_z) n_{0\{n_\perp\}}(\hbar k_z)$$

$$= \frac{1}{2\pi} \sum_{\{n_\perp\}} w_{\{n_\perp\}}(\vec{r}_\perp, \vec{p}_\perp) n_{0\{n_\perp\}}(p_z) \qquad (6)$$

The solution for Eq. (3) can now be found, assuming that the defects are homogenously localized inside the nanowire, so that any dependence of $f(\vec{r}, \vec{p})$ on $z$ can be taken as negligible – i.e. the collision mean field approximation which should be valid, for instance, in a dirty metal regime [22,23] – then, Eq. (3) can be written as

$$\frac{1}{m^*}\vec{p}_\perp \cdot \vec{\nabla}_{r_\perp} f(\vec{r},\vec{p}) + \frac{1}{i\hbar}\left[V_{conf}\left(\vec{r}_\perp + \frac{i\hbar}{2}\vec{\nabla}_{p_\perp}\right) - V_{conf}\left(\vec{r}_\perp - \frac{i\hbar}{2}\vec{\nabla}_{p_\perp}\right)\right] f(\vec{r},\vec{p}) + eE\frac{\partial f(\vec{r},\vec{p})}{\partial p_z}$$
$$= -\left(\frac{\partial f(\vec{r},\vec{p})}{\partial t}\right)_{coll} . \tag{7}$$

In such a model, collisions are important only when some spatial asymmetry arises in the system. An electric field causes such asymmetry, and the solution can be written as a sum of $f_0(\vec{r},\vec{p})$ and a deviation from the equilibrium function due to the electric field $\vec{E}$, which affects only the axial part of the solution,

$$f(\vec{r},\vec{p}) = \frac{1}{2\pi}\sum_{\{n_\perp\}} w_{\{n_\perp\}}(\vec{r}_\perp,\vec{p}_\perp) n_{\{n_\perp\}}(p_z), \tag{8}$$

where $n_{\{n_\perp\}}(p_z) = n_{0\{n_\perp\}}(p_z) + g_{\{n_\perp\}}(p_z)$, with $g_{\{n_\perp\}}(p_z)$ representing a shift in the equilibrium distribution, which can be effectively used to calculate the current density [24]. The basic idea behind Eq. (8) is that the electric field tends to change the electrons occupation numbers, creating an unbalance between the populations of positive and negative $p_z$, which is mediated by collisions, generating a stationary electric current.

For real systems, by considering practical values for the electric field $\vec{E}$, $g_{\{n_\perp\}}(p_z)$ can be taken as a linear response, so that, putting Eq. (8) in (7), we found the following relation

$$\frac{eE}{2\pi}\sum_{\{n_\perp\}} w_{\{n_\perp\}}(\vec{r}_\perp,\vec{p}_\perp)\frac{\partial n_{0\{n_\perp\}}(p_z)}{\partial p_z} = -\frac{1}{2\pi}\sum_{\{n_\perp\}} w_{\{n_\perp\}}(\vec{r}_\perp,\vec{p}_\perp)\left(\frac{\partial n_{\{n_\perp\}}(p_z)}{\partial t}\right)_{coll}, \tag{9}$$

which results in the following simpler form

$$eE\frac{\partial n_{0\{n_\perp\}}(p_z)}{\partial p_z} = -\left(\frac{\partial n_{\{n_\perp\}}(p_z)}{\partial t}\right)_{coll}, \tag{10}$$

and the goal is to find the form of collisional term on the right side of Eq. (10).

Strictly speaking, the effect of collisions is changes the quantum state of electrons. Such changes occur according with a given rate while the electronic current flows, so that

$$\left(\frac{\partial n_{\{n_\perp\}}(p_z)}{\partial t}\right)_{coll} = \sum_{\{n'_\perp\}}\int\frac{dp'_z}{\hbar} W_{\{n_\perp,n'_\perp\}}(p_z,p'_z)\left(n_{\{n'_\perp\}}(p'_z) - n_{\{n_\perp\}}(p_z)\right). \tag{11}$$

where $dp'_z W_{\{n_\perp,n'_\perp\}}(p_z,p'_z)\hbar^{-1}$ is the probability amplitude of an electron in the state with the set of transversal quantum numbers $\{n_\perp\}$, and momentum between $p_z$ and $p_z + dp_z$, be scattered to a state with quantum numbers $\{n'_\perp\}$, and momentum between $p'_z$ and $p'_z + dp'_z$, per unit time. The terms between brackets in Eq. (11) arose from the Fermi Exclusion Principle.

At the conducting band, the scattering can be taken as restrict to the energy shell, and therefore the probability amplitude can be written as

$$W_{\{n_\perp,n'_\perp\}}(p_z,p'_z) = \varpi_{\{n_\perp,n'_\perp\}}(p_z,p'_z)\hbar\delta(\varepsilon - \varepsilon'), \tag{12}$$

where $\varepsilon$ and $\varepsilon'$ are the energies of the initial and final states respectively. Therefore, by the microscopic time symmetry, we have the matrix elements $\varpi_{\{n_\perp,n'_\perp\}}(p'_z,p_z) = \varpi_{\{n_\perp,n'_\perp\}}(p_z,p'_z)$, so that Eq. (11) can be written in the form

$$\left(\frac{\partial n_{\{n_\perp\}}(p_z)}{\partial t}\right)_{coll} = \sum_{\{n'_\perp\}} \frac{m^*}{P} \varpi_{\{n_\perp,n'_\perp\}}(p_z,P)\left(2n_{\{n_\perp\}}(p_z) - n_{\{n'_\perp\}}(P) - n_{\{n'_\perp\}}(-P)\right). \tag{13}$$

The arguments in $P = P_{\{n_\perp,n'_\perp\}}(p_z) = \sqrt{p_z^2 + \hbar^2\left(\kappa_{\{n_\perp\}}^2 - \kappa_{\{n'_\perp\}}^2\right)}$ were intentionally omitted to make the notation simpler. Limits on the sum in Eq. (13) shall be determined by the condition $P^2 \geq 0$.

Knowing that effects of collisions are expected to vanish in average in the absence of electric field, we have

$$\sum_{\{n'_\perp\}} \frac{m^*}{P} \varpi_{\{n_\perp,n'_\perp\}}(p_z,P)\left(2n_{0\{n_\perp\}}(p_z) - 2n_{0\{n'_\perp\}}(P)\right) = 0. \tag{14}$$

In fact, it is easy to show that $n_{0\{n'_\perp\}}(P) = n_{0\{n_\perp\}}(p_z)$ for the Fermi-Dirac distribution, since its argument is the energy $\varepsilon$. Moreover, as $g_{\{n'_\perp\}}(-P) = -g_{\{n'_\perp\}}(P)$, an expression to the collision integral can be finally formulated,

$$\left(\frac{\partial n_{\{n_\perp\}}(p_z)}{\partial t}\right)_{coll} = \left(\sum_{\{n'_\perp\}} \frac{2m^*}{P} \varpi_{\{n_\perp,n'_\perp\}}(p_z,P)\right) g_{\{n_\perp\}}(p_z) = \frac{1}{\tau_{\{n_\perp\}}(p_z)} g_{\{n_\perp\}}(p_z). \tag{15}$$

The function $\tau_{m,n}(p_z)$, as defined in Eq. (15), is the relaxation time for an electron in the state $(m,n,p_z)$, being related to the mean time between collisions.

A simplified equation for the occupation number can be expressed by putting Eq. (15) in the Eq. (10), which gives us

$$eE \frac{\partial n_{0\{n_\perp\}}(p_z)}{\partial p_z} = -\frac{1}{\tau_{\{n_\perp\}}(p_z)} g_{\{n_\perp\}}(p_z), \tag{16}$$

where terms of order $E^2$ were ignored, since these terms are too small for practical applications. Moreover, they would give an even contribution to the final solution, which will be irrelevant to the current computation.

We can find the displacement term $g_{\{n_\perp\}}(p_z)$ by directly solving Eq. (16),

$$g_{\{n_\perp\}}(p_z) = -eE\tau_{\{n_\perp\}}(p_z) \frac{\partial n_{0\{n_\perp\}}(p_z)}{\partial p_z}, \tag{17}$$

so that our next step is to stablish the form of the relaxation time, $\tau_{\{n_\perp\}}(p_z)$, which appears in the solution obtained.

*The Relaxation Time*

In order to obtain the relaxation time, we shall calculate the matrix element $\varpi_{\{n_\perp,n'_\perp\}}(p_z,p'_z)$. By the Fermi's Golden Rule, the transition probability per unit time, $W_{\{k,k'\}}$, between states $\{k\}$ and $\{k'\}$, with respective energies $\varepsilon$ and $\varepsilon'$, assuming that the collisions are mainly due to localized scatter centres, is

$$W_{\{k,k'\}} = \frac{1}{\hbar} \sum_{i=1}^{N_d} \left| U_{\{k,k'\}}(\vec{r}_{0i}) \right|^2 \delta(\varepsilon - \varepsilon'), \tag{18}$$

where $N_d$ is the total number of defects in the structure and $U_{\{k,k'\}}(\vec{r}_{0i}) = \langle k | U(\vec{r} - \vec{r}_{0i}) | k' \rangle$ is the matrix element for transitions due to a single scatter centre located at position $\vec{r}_{0i}$ (supposing that they are all identical).

It is interesting to set the position $\vec{r}_{0i}$ as a generic coordinate, which appoints for any given scatter centre. However, such a system has no symmetry in transversal directions, so that the scattering positions needs to be considered. For general cases, the localization of centres inside de media are stablished, and Eq.(3) results in an integro-differential equation, being solved by suitable computational technics, such as Monte Carlo methods [25,26].

On the other hand, if we know the radial distribution of defects, an average can be taken. By considering such a distribution as homogeneous, for sake of simplicity, we can use the following replacement:

$$\sum_{i=1}^{N_d} \left| U_{\{k,k'\}}(\vec{r}_{0i}) \right|^2 \to \frac{N_d}{A} \int dx_0 dy_0 \left| U_{\{k,k'\}}(x_0, y_0, 0) \right|^2, \tag{19}$$

being $A$ the area of a transversal section of the media under consideration, and $z_0 = 0$ without loss of generality (longitudinal symmetry).

In a simpler approach, the scatter potential should be considered as being spherically symmetrical and sufficiently well localized, so that it can be described by a Dirac's delta function in three-dimensional space

$$U(x - x_0, y - y_0, z) = \gamma \delta(x - x_0) \delta(y - y_0) \delta(z), \tag{20}$$

where $\gamma$ is a constant that gives the potential's intensity, which is localized in the position $(x_0, y_0, 0)$. By considering Eqs. (20) and (19) in Eq. (16) we found the following expression for the transition probability per unit time,

$$W_{\{n_\perp, n'_\perp\}} = \frac{\rho_d}{\hbar} \gamma^2 \int dx_0 dy_0 \left| \psi_{\{n_\perp\}}(x_0, y_0) \psi^*_{\{n'_\perp\}}(x_0, y_0) \right|^2 \delta(\varepsilon - \varepsilon'), \tag{21}$$

being $\rho_d$ the number of defect per volume unit, and the wave functions $\psi_{\{n_\perp\}}(x, y)$ the confinement solution.

By comparing Eq. (21) with Eq. (12), matrix elements $\varpi_{m,m',n,n'}(p_z, p'_z)$ can be identified as

$$\varpi_{\{n_\perp, n'_\perp\}} = \frac{\rho_d}{\hbar^2} \gamma^2 \int dx_0 dy_0 \left| \psi_{\{n_\perp\}}(x_0, y_0) \psi^*_{\{n'_\perp\}}(x_0, y_0) \right|^2, \tag{22}$$

where the momentum variable arguments $(p_z, p'_z)$ are naturally unnecessary. This results in the following expression for the relaxation time in Eq. (15):

$$\tau_{\{n_\perp\}}(p_z) = \left( \sum_{\{n'_\perp\}} \frac{2m^*}{P_{\{n_\perp, n'_\perp\}}(p_z)} \frac{\rho_d \gamma^2}{\hbar^2} \int dx_0 dy_0 \left| \psi_{\{n_\perp\}}(x_0, y_0) \psi^*_{\{n'_\perp\}}(x_0, y_0) \right|^2 \right)^{-1}. \tag{23}$$

The quantity $\rho_d \gamma^2 / \hbar^2$ can be estimated by calculating the relaxation time in a bulk structure. The formulation presented here, in such a case, gives us, for an electron at the Fermi surface,

$$\frac{1}{\tau_F} = \frac{\rho_d}{\hbar^2}\gamma^2\left(4\pi\frac{m^*k_F}{\hbar}\right), \tag{24}$$

which is independent of the set of quantum numbers, since the Fermi surface in a bulk medium is a sphere, so that $\tau_F$ can be taken as a constant of the material.

For $\tau_F \approx 10^{-14} s$, $m^*/m \approx 10^{-2}$, and $k_F \approx 10^7 m^{-1}$, which can be regarded as typical of semiconductor materials, we get, by Eq. (64), $\rho_d\gamma^2\hbar^{-2} \approx 10^4 m^3/s^2$, which can be useful for phenomenological approaches.

## II. SOME RESULTS AND COMMENTS

A nanowire presenting a square transversal section, as shown in Figure 1, can represents a semiconductor device, for instance, upon which an electric field is applied in the $z$ direction in order to generate an electronic current. The conducting electrons are so confined in the transversal directions.

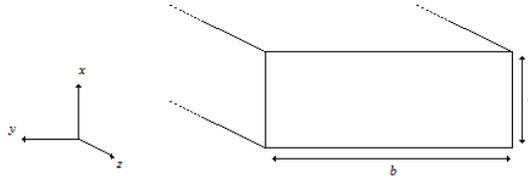

Fig.1. Square transversal section of a nanowire.

The confinement potential is here considered as an infinite wheel, while electrons are free to move in the axial $z$ direction. The defects in this case are homogeneously distributed inside the crystalline structure and the localized spherically symmetric potential given by Eq. (20) shall be used.

The one electron transversal wave function in this case is given by

$$\psi_{n_x,n_y}(x,y) = \frac{2}{\sqrt{ab}}\sin\left(n_x\frac{\pi x}{a}\right)\sin\left(n_y\frac{\pi y}{b}\right), \tag{25}$$

which depends on the set of transversal quantum numbers $\{n_\perp\} = \{n_x, n_y\}$ and the lengths $a$ and $b$ as defined in the Figure 1. We can calculate the relaxation time, Eq. (23), as

$$\tau_{n_x,n_y}(p_z) = \left(\sum_{n'_x,n'_y}\frac{2m^*}{P_{n_x,n_y,n'_x,n'_y}(p_z)}\frac{\rho_d\gamma^2}{\hbar^2}\int dx_0 dy_0 \left|\psi_{n_x,n_y}(x_0,y_0)\psi^*_{n'_x,n'_y}(x_0,y_0)\right|^2\right)^{-1}, \tag{26}$$

where

$$P_{n_x,n'_x,n_y,n'_y}(p_z) = \sqrt{p_z^2 + \hbar^2\left(k_{n_x,n_y}^2 - k_{n'_x,n'_y}^2\right)}, \tag{27}$$

and

$$k_{n_x,n_y}^2 \equiv k_{n_x}^2 + k_{n_y}^2 = \left(n_x\frac{\pi}{a}\right)^2 + \left(n_y\frac{\pi}{b}\right)^2. \tag{28}$$

The wave function shape, Eq. (25), makes the integral in Eq. (26) analytical. Meanwhile, at the Fermi surface, we have

$$p_z = \hbar\sqrt{k_F(a,b)^2 - k_{n_x,n_y}^2}\,, \tag{29}$$

which gives us

$$P_{n_x,n'_x,n_y,n'_y}\left(\hbar\sqrt{k_F(a,b)^2 - k_{n_x,n_y}^2}\right) = \hbar\sqrt{k_F(a,b)^2 - k_{n'_x,n'_y}^2}\,, \tag{30}$$

where $k_F(a,b)$ is the wave number associated to the chemical potential $(\hbar^2/2m^*)k_F(a,b)^2$ which keeps the electronic density constant regardless of dimensions $a$ and $b$. In this context, $k_F$ is defined as the bulk Fermi wavenumber, i.e. $k_F \equiv \lim_{a,b\to\infty} k_F(a,b)$, which is a constant of the media.

Using Eqs. (29) and (30), the relaxation time Eq. (26) can be calculated for the nanowire with a rectangular transversal section. It is interesting to analyse what happens to the values of different relaxation times for different confinement states when the dimensions of the media are changed, to establish whether the constant relaxation time approximation is reasonable or not for a given set of lengths, $a$ and $b$.

For this analysis, two ways were taken: (i) the length $a$ was growth from $5k_F^{-1}$ to $45k_F^{-1}$ keeping $b=5k_F^{-1}$, and so $b$ was grown from $5k_F^{-1}$ to $45k_F^{-1}$, keeping $a=45k_F^{-1}$; (ii) the length $a$ was concomitantly grown with $b$ from $5k_F^{-1}$ to $45k_F^{-1}$. The graphics show the ratio between the mean relaxation time $\langle\tau_F\rangle(a,b)$, at the Fermi surface, and the bulk result, $\tau_F$, given by Eq. (24). The reason for the standard deviation, $\sigma_{\tau_F}(a,b)/\tau_F$, is given as well. The calculation interval is of $0,01k_F^{-1}$.

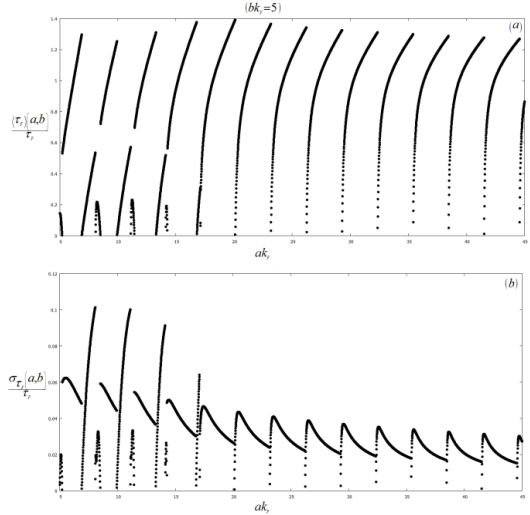

Fig. 2. The graphics for mean relaxation time (a) and standard deviation (b) for the length $a$ going from $5k_F^{-1}$ to $45k_F^{-1}$ keeping $b=5k_F^{-1}$.

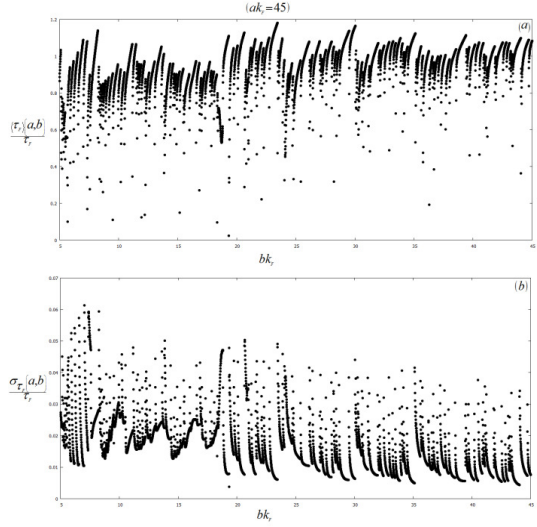

Fig. 3. The graphics for mean relaxation time (a) and standard deviation (b) for the length $b$ going from $5k_F^{-1}$ to $45k_F^{-1}$ keeping $a = 45k_F^{-1}$.

Figures 2 and 3 shows the results for the two aforementioned steps of situation (i). In Figure 2 (a) the relaxation time mean values that were taken over all confinement states are shown as a function of the length $a$, while $b$ is set as constant, which means changing from a nanowire into a nanosheet. Each dotted line in the graphic represents the values of $\langle \tau_F \rangle(a,b)$ for different numbers of confined states allowed inside the nanowire, showing a strong dependence on the dimensions in this case, where confinement is strong.

The anomalies below $20k_F^{-1}$ are related to a change in the number of states due to the chemical potential variation. The graphic in Figure 2 (b) shows that the relative standard deviation can reach significant values in this region, so a constant relaxation time approximation is therefore inadequate.

Figure 3 (a) shows the results for $\langle \tau_F \rangle(a,b)$ with $b$ growing from $5k_F^{-1}$ to $45k_F^{-1}$, keeping $a = 45k_F^{-1}$. A higher accretion rate for the confinement states is seen, with a clear tendency of concentration of points around $\langle \tau_F \rangle(a,b)/\tau_F = 1$ for larger $b$ values, which is expected for the bulk regime. Figure 3 (b) shows that the relative standard deviation is low, so assuming a constant relaxation time can be reasonable.

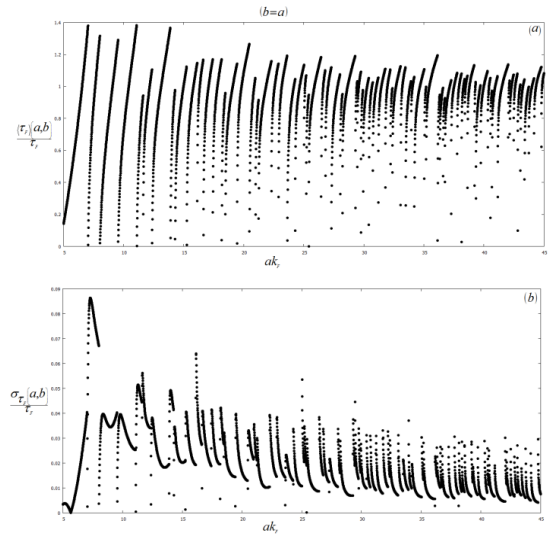

Fig. 4. The graphics for the mean relaxation time (a) and standard deviation (b) for lengths $a$ and $b$ growing concomitantly from $5k_F^{-1}$ to $45k_F^{-1}$.

Finally, Figure 4 shows the results for step (ii). In Figure 4 (a), the relative values for $\langle \tau_F \rangle (a,b)$ are shown for the length $a$ concomitantly growing with $b$ from $5k_F^{-1}$ to $45k_F^{-1}$. The accretion rate grows quadratically due to the increasing in the transversal area. In Figure 4 (b), significant values for standard deviation are seen below $a = 10k_F^{-1}$ (transversal square area $ab < 100k_F^{-2}$).

The results indicate that relaxation time values for different confinement sates become more indistinguishable the more the transversal dimensions are increased, which can be verified by the decrease in $\sigma_{\tau_F}(a,b)/\tau_F$, representing a reduction in the dependency on quantum numbers. This fact justifies the constant relaxation time approach for a bulk structure, where the confinement effects are insignificant.

An example of practical application of the formalism presented here is a chemical sensor. The confinement potential should be affected if the surrounding atmosphere were not chemically inert, changing the relaxation times given in Eq. (23), which causes variations in the conductivity. Moreover, chemical reactions with the nanowire shall make some contributions to defects density, which will be restricted to the surface, originating a distinct contribution to transport relaxation times.

### III. CONCLUSION

We have developed theory for diffusive transport along a nanostructure where confinement effects are important. Specifically, confinement in a direction that is perpendicular to the transport direction. The formalism is based on the Wigner-Boltzmann equation, which is written for a particle quasi-distribution function in the phase space (the phase space formulation of Quantum Mechanics).

When a linear response is expected in the external electric field, and the defects are homogeneously distributed inside the structure, it is found that a relaxation time appears naturally in the theory. This fact results from the assumption that defects distribution do not affect the symmetry of the unperturbed problem in average, making an important contribution only when the external field is applied as a first order perturbation, which breaks this symmetry. The relaxation times are then related to the rate at which the electrons change their quantum states.

The resulting formalism was then applied to a diffusive nanowire and the resulting values for the relaxation time were computed, showing the dispersion of these values according to the confinement strength. It was shown that the constant relaxation time approximation should be reasonable when confinement dimensions are large when compared to the Fermi wavelength of the conduction band.

The present result can serve as a guide for further calculations involving relaxation time approaches. Depending on the confinement potential, diffusive agents and external fields under consideration, this formalism can provide suitable expressions for charge and current densities.

### ACKNOWLEDGEMENT

The author thanks J. d'Albuquerque e Castro and P. F. Farinas for valuable comments and discussions.

### REFERENCES

[1] Silvio R. A.Salinas, "Introduction to Statistical Physics" Springer New York (2001).

[2] Reichl, Linda E. "A Modern Course in Statistical Physics" Wiley-VCH Berlin (2009).

[3] C.K. Zachos, D.B. Fairlie and T.L. Curtright, *World Scientific Series in 20th Century Physics*, **Vol.34** (2005).

[4] C. Jacoboni, R. Brunetti and S. Monastra, *Phys. Rev. B* **68** 125205 (2003).

[5] K. Zalewsnk, *Act Phys Pol B* **34** 3379-3388 (2003).

[6] E. Wigner, *Phys. Rev* **40** 749-759 (1932).

[7] Weyl, H. "Quantenmechanik und Gruppentheorie". Zeitschrift für Physik **46** 1–46 (1927).

[8] Groenewold, H. J. "On the Principles of elementary quantum mechanics". Physica **12** (7) 405–446 (1946).

[9] Moyal, J. E.; Bartlett, M. S. "Quantum mechanics as a statistical theory". Mathematical Proceedings of the Cambridge Philosophical Society **45** 99 (1949).

[10] Jacoboni, C.; Bordone, P. "The Wigner-function approach to non-equilibrium electron transport". Rep. Prog. Phys. **67** 1033–1071 (2004).

[11] M. Nedjalkov, S. Selberherr, I. Dimov, "Stochastic Algorithm for Solving the Wigner-Boltzmann Correction Equation", Numerical Methods and Applications, Lecture Notes in Computer Science **Vol. 6046**, pp 95-102 (2011).

[12] M. Nedjalkov, S. Selberherr, D.K. Ferry, D. Vasileska, P. Dollfus, D. Querlioz, I. Dimov, P. Schwahae, "Physical scales in the Wigner–Boltzmann equation", *Ann Phys* (N Y), **328(C)**: 220–237 (2013).

[13] Yuan, P., Schaefer, L., "Equations of State in a lattice Boltzmann model", Physics of Fluids, **Vol. 18**, (2006).

[14] Harting, J., Chin, J., Maddalena, V., Coveney, P., "Large-scale lattice Boltzmann simulations of complex fluids: advances through the advent of computational Grids", *Philosophical Transactions of the Royal Society A*, **vol. 363**, 1895–1915 (2005).

[15] T. Grasser, H. Kosina, and S. Selberherr, "On the Validity of the Relaxation Time Approximation for Macroscopic Transport Models" Simulation of Semiconductor Processes and Devices **2004** pp. 109-112 (2004).

[16] Quanhua Sun, Chun-Pei Cai, Wei Gao, "On the validity of the Boltzmann-BGK model through relaxation evaluation", Acta Mechanica Sinica, **Vol. 30**, Issue 2, 133-143 (2004).

[17] Yamada, Y.; Tsuchiya, H.; Ogawa, M. "Quantum Transport Simulation of Silicon-Nanowire Transistors Based on Direct Solution Approach of the Wigner Transport Equation". IEEE TRANSACTIONS ON ELECTRON DEVICES, **Vol. 56**, Issue 7, (2009).

[18] Barraud, S; Poiroux, T.; Faynot, O. "A Wigner Function-Based Determinist Method for the Simulation of Quantum Transport in Silicon Nanowire Transistors". IEEE (2011).

[19] Demeio, L.; Barletti, L.; Bordone, P.; Jacoboni, C. "WIGNER FUNCTION FOR MULTIBAND TRANSPORT IN SEMICONDUCTORS" Transport Theory and Statistical Physics **32** 3-4 (2003).

[20] M. A. Reed, J. N. Randall, R. J. Aggarwal, R. J. Matyi, T. M. Moore, and A. E. Wetsel, "Observation of discrete electronic states in a zero-dimensional semiconductor nanostructure", Phys. Rev. Lett. **60**, 535 (1988).

[21] Paulo F. Farinas, Gilmar E. Marques, and Nelson Studart, "Subband mixing in resonant magnetotunneling through double-barrier semiconductor nanostructures", Journal of Applied Physics **79**, 8475 (1996).

[22] Lee, P. A.; Ramakrishnan, T. V. "Disordered electronic systems". Rev Mod Phys **57** 287-337 (1985).

[23] Beenakker, C. W. J.; van Houten, H. "Quantum Transport in Semiconductor Nanostructures". Solid State Physics, **Vol. 44**, (1991).

[24] A. R. Fernandes Nt, J. A. Otálora, P. Vargas, and J. d'Albuquerque e Castro, "Oscillations in the spatial distribution of current in nanotubes and nanowires", Journal of Applied Physics **110**, 093720 (2011).

[25] Damien Querlioz, Jérôme Saint-Martin, Philippe Dollfus, "Implementation of the Wigner-Boltzmann transport equation within particle Monte Carlo simulation" Journal of Computational Electronics, **Vol. 9**, Issue 3-4, pp 224-231 (2010).

[26] J.M. Sellier, I. Dimova, "The Wigner-Boltzmann Monte Carlo Method Applied to Electron Transport in the Presence of a Single Dopant" Computer Physics Communications, **Vol. 185**, Issue 10, 2427–2435 (2014).